\documentclass[english]{article}
\usepackage{a4wide}
\usepackage{float}
\usepackage{babel}

\newtheorem{theorem}{Theorem}
\newtheorem{algorithm}[theorem]{Algorithm}

\begin{document}

\title{A Monte Carlo algorithm for efficient large matrix inversion}

\author{L. A. Garc\'{\i}a-Cort\'{e}s %
\footnote{Dpto. de Mejora Gen\'{e}tica, CIT-INIA, Carretera de La Coruña, Km.
7,5 Madrid, E-28040, Spain. e-mail: garcia@inia.es%
} %
\footnote{Dpto. de Producción Animal. ETSI Agrónomos. Universidad Politécnica
de Madrid%
} and C. Cabrillo %
\footnote{Instituto de Estructura de la Materia, Serrano 123, Madrid, E-28006,
Spain. e-mail: ccabrilo@foton0.iem.csic.es%
}}

\maketitle
\begin{abstract}
This paper introduces a new Monte Carlo algorithm to invert large
matrices. It is based on simultaneous coupled draws from two random
vectors whose covariance is the required inverse. It can be considered
a generalization of a previously reported algorithm for hermitian
matrices inversion based in only one draw. The use of two draws allows
the inversion on non-hermitian matrices. Both the conditions for convergence
and the rate of convergence are similar to the Gauss-Seidel algorithm.
Results on two examples are presented, a real non-symmetric matrix
related to quantitative genetics and a complex non-hermitian matrix
relevant for physicists. Compared with other Monte Carlo algorithms
it reveals a large reduction of the processing time showing eight
times faster processing in the examples studied. 
\end{abstract}

\section{Introduction}

The computation of the inverse of a matrix or some of its elements
is one of the main topics in numerical analysis. Large sparse matrices
can be usually inverted from its factors obtained by using sparse
matrix techniques \cite{george}. Although it can be alleviated by
reordering rows and columns, these techniques suffer from the fill-in
required by the LU factorization and they usually result in algorithms
demanding large random access memory (RAM) values.

Monte Carlo algorithms are used to estimate empirically the expectation
of random variables. Using them to invert matrices consist basically
in obtaining and averaging several realized values of a variable whose
expectation is the required inverse. Although Monte Carlo algorithms
present a stochastic error, they can be used to obtain estimates of
the inverse when the matrix is too large or complicate to be inverted
by using conventional or sparse algorithms. These algorithms have
high parallel efficiency, they do not demand a great amount of RAM
and the processing time is proportional to the number of nonzero elements
of the matrix to be inverted. In general, they can be very efficient
when rough estimates of the inverse matrix are required.

The most common Monte Carlo algorithms to invert matrices are based
on finite discrete Markov chains \cite{dimov}. In these algorithms
several random trajectories are calculated to solve a linear system
of equations for each matrix column. Alternatively, there are Monte
Carlo methods based on random variables whose expectation is the whole
inverse instead one of its columns. Typically the rational is to consider
the sought for inverse matrix $\mathbf{C}^{-1}$ as the covariance
matrix of a normally distributed random vector, \begin{equation}
\mathbf{z}\sim N\left(0,\mathbf{C}^{-1}\right)\,.\label{eq1}\end{equation}

\noindent After generating by any Monte Carlo algorithm sampled values
of $\mathbf{z}$ (something for which $\mathbf{C}^{-1}$ is not needed
but just $\mathbf{C}$), the inverse matrix is obtained by an estimation
of the variances and covariances of the sample.

Recently the Gibbs sampler \cite{gelfand:90a,geman:84} algorithm
has been proposed to draw $\mathbf{z}$ values from (\ref{eq1}) using
the elements of $\mathbf{C}$ \cite{garcia-cortes:95itea,harville:99}.
The procedure consists in updating successively the elements of $\mathbf{z}$
for several cycles by using the formula \begin{equation}
z_{i}^{\left(k\right)}=\phi_{i}^{\left(k\right)}\frac{1}{\sqrt{c_{ii}}}-\frac{1}{c_{ii}}\sum_{j=1}^{i-1}z_{j}^{\left(k\right)}c_{ij}-\frac{1}{c_{ii}}\sum_{j=i+1}^{n}z_{j}^{\left(k-1\right)}c_{ij}\label{eq2}\end{equation}

\noindent where the subscript $i$ means the $i^{th}$ element of
$\mathbf{z}$, the superscript $(k)$ means the $k^{th}$ cycle and
$\phi$ is an independent random noise drawn from a standard normal
distribution. After a given period of convergence usually called the
burn-in period, $\mathbf{z}^{(k)}$ will satisfy at each cycle $E\left(\mathbf{z}^{(k)}\mathbf{z}^{\dagger(k)}\right)=\mathbf{C}^{-1}$
or $E\left(\mathbf{z}^{\dagger(k)}\mathbf{Qz}^{(k)}\right)=tr\mathbf{QC}^{-1}$
(the $\dagger$ denotes hermitian conjugate, i.e., transposition and
complex conjugation). These expectations will provide the elements
of the inverse or linear functions of them respectively. Hereafter,
we will denote this algorithm by \emph{GS}. Although \emph{GS} does
not provide an independent but a serially correlated set of samples,
the convergence is fast and, contrary to Metropolis-like algorithms,
it profits of all the draws. 

The obvious disadvantage of the algorithms based in Eq. (\ref{eq1})
is a rather narrow range of applicability because of the necessary
condition of $\mathbf{C}$ being positive definite. For many statistical
applications this is not a problem \cite{harville:99} but it is a
serious drawback in Physics applications (for instance in Lattice
QCD \cite{kogut:83}). The problem can be bypassed by inverting $\mathbf{C}^{\dagger}\mathbf{C}$
and applying the result to $\mathbf{C}$ but this procedure seriously
reduces the efficiency. Rather surprisingly this is not the case for
the \emph{GS} algorithm\emph{.} Once recourse is made to the Gibbs
sampler to implement the covariance matrix rational, the convergence
of the algorithm, at least if it is explicitly written as in Eq. (\ref{eq2}),
is governed by the sampling method and not by the implications of
(\ref{eq1}). As we will see, the convergence conditions are then
more flexible but still$\mathbf{C}$must be symmetric (or hermitian
if defined over the complex numbers).

A different approach usually known as stochastic estimation (\emph{SE}
from now on), avoid this problem \cite{gusken:99} by recourse to
the repeated solution of linear systems like in the discrete Markov
Chain methods mentioned above but this time the number of systems
to be solved does not depend on the rank of the matrix to be inverted.
The rational is to solve a series of linear systems $\mathbf{Cv}=\phi$
where $\phi$ is drawn from a random variable satisfying $E\left(\phi\right)=0$
and $E\left(\phi\phi^{\dagger}\right)=\mathbf{I}$. The inverse is
then given by taking the ensemble average $E\left(\mathbf{v}\phi^{\dagger}\right)=\mathbf{C}^{-1}$.
Rather surprisingly it was not recognized until recently that this
method do not rely on the gaussianity of $\phi$. Thus, Dong and Liu
\cite{dong:94} and, independently, Garc\'{\i}a-Cort\'{e}s \cite{garcia-cortes:94guelph}
have shown that the efficiency of the method improves substantially
when is drawn as $\phi_{i}=2B-1$, where $B$ is a Bernoulli random
variable (a $Z^{2}$ noise).

When applicable, however, it is expected that \emph{GS} will be more
efficient than \emph{SE}. This comes from the fact that \emph{GS}
mimics Gauss-Seidel algorithm but including noise at each cycle \cite{thompson:95}.
The processing time to obtain each realized value of $\mathbf{zz}^{\dagger}$
or $\mathbf{z}^{\dagger}\mathbf{Qz}$ is then slightly greater than
the processing time of a Gauss-Seidel iteration, while the processing
time to obtain a \emph{SE} realized value is equivalent to the processing
time needed to solve a linear system.

In this paper we introduce a new ''noisy'' Gauss-Seidel method which
include as a particular case the \emph{GS} algorithm. Our algorithm,
that we will call it the correlated chains sampling algorithm (\emph{CC}
from now on), is applicable to nondefinite positive matrices and profits
of the larger efficiency of the $Z^{2}$ noise. The power of the algorithm
is assessed with two different non-hermitian matrix, a real one used
in genetic improvement in animal breeding and a complex one typical
of particle Physics. Numerical tests show how our proposal can be
around an order of magnitude faster than \emph{SE}. The sufficient
and necessary conditions for the convergence of the method are also
given.

\section{The correlated chains sampling algorithm}

The \emph{CC} algorithm introduced in this section is based on the
simultaneous update of a couple of random vectors $\mathbf{z}$ and
$\mathbf{w}$ such as, after convergence of the algorithm, $E\left(\mathbf{zw}^{\dagger}\right)=\mathbf{C}^{-1}$
being $\mathbf{C}$ the matrix to be inverted. The algorithm updates
simultaneously each element of $\mathbf{z}$ and $\mathbf{w}$ using
\begin{equation}
z_{i}^{\left(k\right)}=\phi_{i}^{(k)}\frac{1}{\sqrt{c_{ii}}}-\frac{1}{c_{ii}}\sum_{j=1}^{i-1}z_{j}^{\left(k\right)}c_{ij}-\frac{1}{c_{ii}}\sum_{j=i+1}^{n}z_{j}^{\left(k-1\right)}c_{ij}\label{eq4}\end{equation}
\begin{equation}
w_{i}^{\left(k\right)}=\phi_{i}^{(k)}\frac{1}{\sqrt{c_{ii}^{*}}}-\frac{1}{c_{ii}^{*}}\sum_{j=1}^{i-1}w_{j}^{\left(k\right)}c_{ji}^{*}-\frac{1}{c_{ii}^{*}}\sum_{j=i+1}^{n}w_{j}^{\left(k-1\right)}c_{ji}^{*}\label{eq5}\end{equation}

\noindent where $\Phi^{(k)}$ is a set of independent noise vectors
such as

\noindent \begin{equation}
E\left(\Phi^{(k)}\right)=0\label{eq5a}\end{equation}
 and \begin{equation}
E\left(\Phi^{(k)}\Phi^{(l)\dagger}\right)=\mathbf{I}\delta_{k,l}.\label{eq5b}\end{equation}
 Note that the noise term in (\ref{eq4}) and in (\ref{eq5}) are
both the same.

After discarding $N$ cycles of a total of $M$ as a period of convergence
or burn-in, the Monte Carlo estimation of the inverse can be obtained
from, \begin{equation}
\mathbf{C}^{-1}\simeq\frac{1}{M-N}\sum_{k=N+1}^{M}\mathbf{z}^{(k)}\mathbf{w}^{(k)\dagger}\,.\label{1/C}\end{equation}

Expression $E\left(\mathbf{z}^{\dagger}\mathbf{Qz}\right)=tr\mathbf{QC}^{-1}$
also holds, and its Monte Carlo estimation via the \emph{CC} algorithm
is given by \begin{equation}
tr\mathbf{QC}^{-1}\simeq\frac{1}{M-N}\sum_{k=N+1}^{M}\mathbf{z}^{\left(k\right)\dagger}\mathbf{Qw}^{\left(k\right)}\label{tr}\end{equation}

\subsection{Demonstration}

Let us start by rewriting Eqs.\ (\ref{eq4}) and (\ref{eq5}) in
matrix form, \begin{equation}
\mathbf{z}^{\left(k\right)}=\frac{1}{\sqrt{\mathbf{D}}}\Phi^{\left(k\right)}-\frac{1}{\mathbf{D}}\mathbf{Lz}^{\left(k\right)}-\frac{1}{\mathbf{D}}\mathbf{Uz}^{\left(k-1\right)}\,,\label{mform1}\end{equation}
\begin{equation}
\mathbf{w}^{\left(k\right)}=\frac{1}{\sqrt{\mathbf{D}^{\dagger}}}\Phi^{\left(k\right)}-\frac{1}{\mathbf{D}^{\dagger}}\mathbf{U}^{\dagger}\mathbf{w}^{\left(k\right)}-\frac{1}{\mathbf{D}^{\dagger}}\mathbf{L}^{\dagger}\mathbf{w}^{\left(k-1\right)}\,,\label{mform2}\end{equation}

\noindent where $\mathbf{L}$, $\mathbf{D}$ and $\mathbf{U}$ are
the lower triangle, the diagonal and the upper triangle of $\mathbf{C}$
respectively, so that $\mathbf{L}+\mathbf{D}+\mathbf{U}=\mathbf{C}$.
Solving for $\mathbf{z}^{\left(k\right)}$ and $\mathbf{w}^{\left(k\right)\dagger}$
from Eqs.(\ref{mform1}) and (\ref{mform2}) we have, \begin{equation}
\mathbf{z}^{\left(k\right)}=\frac{1}{(\mathbf{D}+\mathbf{L})}\sqrt{\mathbf{D}}\,\Phi^{\left(k\right)}-\frac{1}{\mathbf{D}+\mathbf{L}}\mathbf{Uz}^{\left(k-1\right)}\,,\label{zk}\end{equation}
\begin{equation}
\mathbf{w}^{\left(k\right)\dagger}=\Phi^{\left(k\right)\dagger}\sqrt{\mathbf{D}}\frac{1}{(\mathbf{D}+\mathbf{U})}-\mathbf{w}^{\left(k-1\right)\dagger}\mathbf{L}\frac{1}{\mathbf{D}+\mathbf{U}}\,.\label{wk}\end{equation}

\noindent After running the algorithm during $n$ updating cycles
(\ref{zk}) and (\ref{wk}) leads to, \begin{eqnarray}
\mathbf{z}^{(n)} & = & \sum_{k=0}^{n}(-\mathbf{T})^{k}\,\Theta^{(n-k)}\,,\label{zn}\\
\mathbf{w}^{(n)\dagger} & = & \sum_{k=0}^{n}\Gamma^{(n-k)\dagger}\,(-\mathbf{S})^{k}\,,\label{wn}\end{eqnarray}

\noindent where for the ease of the notation we have defined, \begin{eqnarray}
\mathbf{T} & = & \frac{1}{\mathbf{D}+\mathbf{L}}\mathbf{U}\,,\label{T}\\
\mathbf{S} & = & \mathbf{L}\frac{1}{\mathbf{D}+\mathbf{U}}\,,\label{S}\\
\Theta^{(k)} & = & \frac{1}{(\mathbf{D}+\mathbf{L})}\sqrt{\mathbf{D}}\,\Phi^{(k)}\,,\label{Theta}\\
\Gamma^{(k)} & = & \frac{1}{(\mathbf{D}+\mathbf{U})^{\dagger}}\sqrt{\mathbf{D}^{\dagger}}\,\Phi^{(k)}\,,\label{Gamma}\end{eqnarray}
 for all $k$ except for $k=0$ for which, again for the ease of notation,
$\Theta^{(0)}$ and $\Gamma^{(0)}$ denote just the initial vectors
$\mathbf{z}^{(0)}$ and $\mathbf{w}^{(0)}$ respectively. For a bounded
noise such us the $Z^{2}$ noise, \textbf{$\mathbf{z}^{(n)}$} and
$\mathbf{w}^{(n)}$are bounded by\begin{eqnarray*}
\left\Vert \mathbf{z}^{(n)}\right\Vert \leq\left\Vert \sum_{k=0}^{n}(-\mathbf{T})^{k}\right\Vert \left\Vert \frac{1}{(\mathbf{D}+\mathbf{L})}\sqrt{\mathbf{D}}\right\Vert B\leq\left\Vert \frac{1}{(\mathbf{D}+\mathbf{L})}\sqrt{\mathbf{D}}\right\Vert B\,\sum_{k=0}^{n}\left\Vert \mathbf{T}^{k}\right\Vert \,,\\
\left\Vert \mathbf{w}^{(n)}\right\Vert \leq\left\Vert \sum_{k=0}^{n}(-\mathbf{S})^{k}\right\Vert \left\Vert \frac{1}{(\mathbf{D}+\mathbf{U})}\sqrt{\mathbf{D}}\right\Vert B\leq\left\Vert \frac{1}{(\mathbf{D}+\mathbf{U})}\sqrt{\mathbf{D}}\right\Vert B\,\sum_{k=0}^{n}\left\Vert \mathbf{S}^{k}\right\Vert \,,\end{eqnarray*}
 where $B$ is the upper bound of the noise $\Phi$ and therefore
the absolute convergence of the alternating power series $\mathbf{T}$
and $\mathbf{S}$ warrants the convergence of the algorithm. In fact,
this is a necessary and sufficient condition for the convergence of
the algorithm for any possible value of the noise chain. To see this
notice that the worst case corresponds to a noise such as the components
of $\mathbf{T}^{k}$ and $\mathbf{S}^{k}$ with the maximum absolute
values attains positive signs. But the maximum of the absolute values
of the components of a matrix is a norm. The convergence of these
components, which warrants the convergence of any other of the components,
corresponds therefore to the absolute convergence of the series under
such a norm and all the norms are equivalent as far as the convergence
of matrix series is concerned.

Given the convergence of the alternating power series in $\mathbf{T}$
and $\mathbf{S}$ we have finite $\mathbf{z}^{(\infty)}$ and $\mathbf{w}^{(\infty)}$
and for a given integer $N$, the $\mathbf{z}^{(n)}$ and $\mathbf{w}^{(n)}$
corresponding to $n\geq N$ can be split in series of $N$ terms plus
and arbitrarily small remainder for a sufficiently large $N$,\begin{eqnarray*}
\mathbf{z}^{(n)} & = & \sum_{k=0}^{N-1}(-\mathbf{T})^{k}\,\Theta^{(n-k)}+O(\left\Vert \mathbf{T}\right\Vert ^{N+1})\,,\\
\mathbf{w}^{(n)\dagger} & = & \sum_{k=0}^{N-1}\Gamma^{(n-k)\dagger}\,(-\mathbf{S})^{k}+O(\left\Vert \mathbf{S}\right\Vert ^{N+1})\,.\end{eqnarray*}

\noindent The sample average of the $\mathbf{zw}^{\dagger}$ product
discarding $N$ burn-in cycles of a total of $M\gg N$, again retaining
explicitly only terms up to order $N$, gives \begin{eqnarray}
\langle\mathbf{zw}^{\dagger}\rangle=\frac{1}{M-N}\sum_{n=N}^{M}\:\sum_{k=0}^{N-1}\sum_{j=0}^{N-1}(-1)^{k+j}\mathbf{T}^{k}\;\Theta^{(n-k)}\Gamma^{(n-j)\dagger}\;\mathbf{S}^{j} & +\nonumber \\
O(\left\Vert \mathbf{T}\right\Vert ^{N+1},\,\left\Vert \mathbf{S}\right\Vert ^{N+1}) & =\nonumber \\
\sum_{k=0}^{N-1}\sum_{j=0}^{N-1}(-1)^{k+j}\mathbf{T}^{k}\;\frac{1}{M-N}\sum_{n=N}^{M}\Theta^{(n-k)}\Gamma^{(n-j)\dagger}\;\mathbf{S}^{j} & +\nonumber \\
O(\left\Vert \mathbf{T}\right\Vert ^{N+1},\,\left\Vert \mathbf{S}\right\Vert ^{N+1})\,,\label{saverage}\end{eqnarray}
 Taking the limit of large $M$, Eq. (\ref{saverage}) yields,

\noindent \begin{eqnarray}
\lim_{M\rightarrow\infty}\langle\mathbf{zw}^{\dagger}\rangle=E(\mathbf{zw}^{\dagger}) & =\nonumber \\
\sum_{k=0}^{N-1}\sum_{j=0}^{N-1}(-1)^{k+j}\mathbf{T}^{k}E\left(\Theta^{(n-k)}\Gamma^{(n-j)\dagger}\right)\mathbf{S}^{j}+O(\left\Vert \mathbf{T}\right\Vert ^{N+1},\,\left\Vert \mathbf{S}\right\Vert ^{N+1}) & =\nonumber \\
\sum_{k=0}^{N-1}\sum_{j=0}^{N-1}(-1)^{k+j}\mathbf{T}^{k}\frac{1}{\mathbf{D}+\mathbf{L}}\sqrt{\mathbf{D}}\;\mathbf{I}\delta_{k,j}\sqrt{\mathbf{D}}\frac{1}{\mathbf{D}+\mathbf{U}}\mathbf{S}^{j}+O(\left\Vert \mathbf{T}\right\Vert ^{N+1},\,\left\Vert \mathbf{S}\right\Vert ^{N+1}) & =\nonumber \\
\sum_{k=0}^{N-1}\mathbf{T}^{k}\frac{1}{\mathbf{D}+\mathbf{L}}\mathbf{D}\frac{1}{\mathbf{D}+\mathbf{U}}\mathbf{S}^{k}+O(\left\Vert \mathbf{T}\right\Vert ^{N+1},\,\left\Vert \mathbf{S}\right\Vert ^{N+1})\,.\label{series}\end{eqnarray}

\noindent In the limit of large $N$, Eq. (\ref{series}) reduces
to\begin{equation}
\lim_{N\rightarrow\infty}E(\mathbf{zw}^{\dagger})=\sum_{k=0}^{\infty}\mathbf{T}^{k}\frac{1}{\mathbf{D}+\mathbf{L}}\mathbf{D}\frac{1}{\mathbf{D}+\mathbf{U}}\mathbf{S}^{k}\,.\label{finalSeries}\end{equation}

Equation (\ref{finalSeries}) represents the stationary value of the
$\mathbf{zw}^{\dagger}$ average. That the series in (\ref{finalSeries})
approach $\mathbf{C}^{-1}$, when they converge, is easily verified
by iteration of the following recursive formula, \begin{equation}
\frac{1}{\mathbf{C}}=\frac{1}{\mathbf{D}+\mathbf{L}}\mathbf{D}\frac{1}{\mathbf{D}+\mathbf{U}}+\mathbf{T}\frac{1}{\mathbf{C}}\mathbf{S}\,,\label{recursive}\end{equation}

\noindent which can be derived from the identity, \[
\left(\mathbf{C}-\mathbf{U}\right)\frac{1}{\mathbf{C}}\left(\mathbf{C}-\mathbf{L}\right)=\mathbf{C}-\left(\mathbf{U}+\mathbf{L}\right)+\mathbf{U}\frac{1}{\mathbf{C}}\mathbf{L}\,,\]

\noindent since for the particular case in which $\mathbf{C}=\mathbf{D}+\mathbf{U}+\mathbf{L}$
it reduces to, \[
\left(\mathbf{D}+\mathbf{L}\right)\frac{1}{\mathbf{C}}\left(\mathbf{D}+\mathbf{U}\right)=\mathbf{D}+\mathbf{U}\frac{1}{\mathbf{C}}\mathbf{L}\,.\]

\noindent Making use of the definitions (\ref{T}) and (\ref{S})
formula (\ref{recursive}) follows trivially.

\subsection{Convergence analysis}

As shown in the previous section, the convergence of the algorithm
is determined by the absolute convergence of the series,\begin{eqnarray}
\sum_{k=0}^{\infty}(-\mathbf{T})^{k}\,,\label{Tseries}\\
\sum_{k=0}^{\infty}(-\mathbf{S})^{k}\,.\label{Sseries}\end{eqnarray}
 We therefore center in searching for the necessary and sufficient
conditions for the convergence of the series (\ref{Tseries}) and
(\ref{Sseries}). Restricting the analysis to the series (\ref{Tseries})
as the same rational applies to (\ref{Sseries}), first notice that
the absolute convergence of (\ref{Tseries}) implies that the spectral
radius of $\mathbf{T}$, i.e., $\lim_{k\rightarrow\infty}\left\Vert \mathbf{T}^{k}\right\Vert ^{1/k}\equiv sp(\mathbf{T})$
\cite{Bau:85}, is strictly below 1. This comes as a consequence of
the Cauchy root convergence test \cite{Arfen:85} which implies $\left\Vert \mathbf{T}^{m}\right\Vert ^{1/m}<1$
for some sufficiently large $m$ if (\ref{Tseries}) converges. Conversely
if $sp(\textrm{T})<1$, then $\left\Vert \mathbf{T}^{m}\right\Vert ^{1/m}<1$
for some large enough $m$ which in turn implies $\left\Vert \mathbf{T}^{m}\right\Vert <1$.
We now split the series of the absolute values of (\ref{Tseries})
in a finite part with the terms up to $m-1$ and the rest, so that,

\[
\sum_{k=0}^{\infty}\left\Vert \mathbf{T}^{k}\right\Vert =\sum_{k=0}^{m-1}\left\Vert \mathbf{T}^{k}\right\Vert +\sum_{k=0}^{\infty}\left\Vert \mathbf{T}^{m+k}\right\Vert \leq\sum_{k=0}^{m-1}\left\Vert \mathbf{T}^{k}\right\Vert +\sum_{k=0}^{\infty}\left\Vert \mathbf{T}^{m}\right\Vert ^{k}\,.\]
 Given that $\left\Vert \mathbf{T}^{m}\right\Vert <1$the last term
converges to $1/(1-\left\Vert \mathbf{T}\right\Vert )$ and therefore
(\ref{Tseries}) converges absolutely. Applying the same rational
to (\ref{Sseries}) we have then, that the algorithm converges in
the sense that both $\mathbf{z}^{(\infty)}$ and $\mathbf{w}^{(\infty)}$
are finite for any possible drawn values of the noise $\Phi$, if
and only if $sp(\mathbf{T})<1$ and $sp(\mathbf{S})<1$ . In strict
sense, these are the necessary and sufficient conditions for the convergence
of Eqs. (\ref{zk}) and (\ref{wk}). The final algorithm is implemented
using (\ref{mform1}) and (\ref{mform2}) and therefore also the non-singularity
of $\mathbf{D}$is needed.

\subsection{The burn-in period and Monte Carlo error}

As usual in this kind of algorithms, the efficiency of the code is
increased by discarding a number of iterations, $N$, called burn-in
period, which are too influenced by the initial values. Given that
the \emph{CC} path does not converge to a deterministic value but
to a random variable an estimation of $N$ is not as easy as in a
deterministic iterative procedures (for instance, the Gauss-Seidel
algorithm). To overcome the difficulties associated to the random
nature of the algorithm we will use the so called coupling method
\cite{johnson:96}. It consists in running a couple of paths for both
$\mathbf{z}$ and $\mathbf{w}^{\dagger}$ with different initial values
but with the same random numbers for each couple. When the difference
between paths of a couple reach a given tolerance, the burn-in is
assumed to be finished. From (\ref{mform1}) the difference between
two paths $\mathbf{z}_{1}$ and $\mathbf{z}_{2}$ with different initial
values but with the same $\Phi^{(k)}$ results in, \begin{equation}
\mathbf{z}_{2}^{\left(k\right)}-\mathbf{z}_{1}^{\left(k\right)}=-\frac{1}{\mathbf{D}}\mathbf{L}\left(\mathbf{z}_{2}^{\left(k\right)}-\mathbf{z}_{1}^{\left(k\right)}\right)-\frac{1}{\mathbf{D}}\mathbf{U}\left(\mathbf{z}_{2}^{\left(k-1\right)}-\mathbf{z}_{2}^{\left(k-1\right)}\right)\,.\label{eq12}\end{equation}
 But Eq. (\ref{eq12}) is nothing else that the formula corresponding
to a Gauss-Seidel iterative algorithm \cite{young:71} for the solution
of the linear system, \[
\mathbf{C}\left(\mathbf{z}_{2}-\mathbf{z}_{1}\right)=0\,,\]
 whose solution is $\mathbf{z}_{2}-\mathbf{z}_{1}=0$, for a non-singular
$\mathbf{C}$. Not surprisingly, the condition for the convergence
of the Gauss-Seidel algorithm represented by (\ref{eq12}) is, $sp(\left(\mathbf{D}+\mathbf{L}\right)^{-1}\mathbf{U})<1$
, that is, $sp(\mathbf{T})<1$. As expected, the same rational applied
to Eqs. (\ref{mform2}) leads to a Gauss-Seidel algorithm for $\mathbf{C}^{\dagger}(\mathbf{w}_{2}-\mathbf{w}_{1})=0$
with a convergence condition $sp(\left(\mathbf{D}^{\dagger}+\mathbf{U}^{\dagger}\right)^{-1}\mathbf{L}^{\dagger})<1$,
that is, $sp(\mathbf{S})<1$ . These two Gauss-Seidel procedures can
be useful to analyze the convergence rate of the \emph{CC} algorithm
in a given instance.

It is also relevant to calculate the Monte Carlo error of the expectations
(\ref{1/C}). The Monte Carlo error is Gaussian because of the central
limit theorem, but the realized values provided by the \emph{CC} algorithm
are serially correlated and its Monte Carlo variance is bigger than
the Monte Carlo variance expected from an independent set of $M-N$
samples. To calculate the Monte Carlo error instead of $M-N$ samples,
we used a higher effective number of samples as proposed in \cite{geyer:92}.

\subsection{Algorithm}

\emph{Algorithm 1} shows the correlated chains algorithm including
the determination of the burn-in period ($N$). During the burn-in
period, four chains are computed until the paths converge to a given
tolerance, typically a number small enough to avoid the dependency
of the chains on the starting values. In order to simplify the algorithm
outlined here, we consider a fixed chain length ($M$), i.e., we are
assuming that the $M-N$ cycles after burn-in are enough to obtain
accurate estimates of the elements of the inverse. Nevertheless, in
the numerical tests presented in this paper, we have included the
determination of $M-N$ by inserting the calculation of the effective
number of cycles during the iteration process.

\begin{algorithm}

\textsc{Correlated chains algorithm including the determination of the burn-in period}

Given \(n\), \(\mathbf{Q}\), \(\mathbf{C}\), \(B\), \(tol\)

1. Set arbitrary starting values for \(\mathbf{z}\), \(\mathbf{z}^{\ast }\), \(\mathbf{w}^{\prime }\) and \(\mathbf{w}^{\prime \ast }\), for instance \(z_{i}\mathbf{=}0\), \(z_{i}^{\ast }\mathbf{=}i\), \(w_{i}\mathbf{=}0\) and \(w_{i}^{\ast }\mathbf{=}i\).

2. Sample \(\mathbf{\phi }\) as a vector containing independent draws according with definition (\ref{eq5a}) and (\ref{eq5b})

3. Update \(\mathbf{z}\), \(\mathbf{z}^{\ast }\), \(\mathbf{w}^{\prime }\) and \(\mathbf{w}^{\prime \ast }\) by using equations (\ref{eq4}) and (\ref{eq5})

4. \(p\leftarrow p+1\)

5. Go to step 2 until \(\mathbf{z}^{\prime }\mathbf{z}^{\ast }<tol\)\ and \(\mathbf{w}^{\prime }\mathbf{w}^{\ast }<tol\)

6. Sample \(\mathbf{\phi }\) as a vector containing independent draws according with definition (\ref{eq5a}) and (\ref{eq5b})

7. Update \(\mathbf{z}\)\ and \(\mathbf{w}\) by using equations (\ref{eq4}) and (\ref{eq5})

8. Accumulate \(\mathbf{z}^{\prime }\mathbf{Qw}\) in \(\mathbf{s}\)

9. Go to step 6 to compute the next round of iteration (\(B-p\) times)

10. Set the final estimate: \(tr\mathbf{QC}^{-1}\leftarrow \mathbf{s}/\left( B-p\right) \)

\end{algorithm}

\section{Numerical tests}

To check the computational efficiency of the proposed algorithm we
will compare its performance with respect to the efficiency of the
\emph{SE} algorithm which has shown itself as a efficient inversion
method \cite{dong:94}. We will make use of two different examples
from two disparate fields: genetic improvement in animal breeding
and quantum field theory in the lattice.

\subsection{Example 1. Wu and Schaeffer's real asymmetric matrix}

The Henderson's mixed model equations \cite{henderson:84a} are routinely
used in animal breeding to evaluate the candidates to the artificial
selection in livestock populations. This method takes into account
both the performance records of the animals in a given population
and the pedigree relationships between animals. In animal breeding,
the quality of the estimations provided by a given model corresponds
to the diagonal elements of the inverse of a coefficient matrix. For
instance, in the case of the simplest model, these diagonal elements,
known as the prediction error variances, take the form \begin{equation}
diag\left[\left(\begin{array}{cc}
\mathbf{X}^{\dagger}\mathbf{X} & \mathbf{X}^{\dagger}\\
\mathbf{X} & \mathbf{I}+\mathbf{A}^{-1}\frac{\sigma_{e}^{2}}{\sigma_{a}^{2}}\end{array}\right)^{-1}\sigma_{e}^{2}\right]\label{eq13}\end{equation}
 where $\mathbf{X}$ is an incidence matrix mapping animals into herds.
Its nonzero elements are $x_{ij}=1$, which indicates that animal
$i$ was recorded in herd $j$. $\mathbf{I}$ is the identity matrix
with order equal to the number of animals in the population. Variances
$\sigma_{e}^{2}$ and $\sigma_{a}^{2}$ are assumed to be known and
correspond to the additive genetic variance and the residual variance.
$\mathbf{A}$ is known as the numerator relationship matrix and it
maps the genetic relationships between animals. For instance, $a_{ij}=0.25$
if animals $i$ and $j$ are half-sibs, $a_{ij}=0.5$ if animals $i$
and $j$ are parent-progeny related, $a_{ii}=1$ if the animal $i$
is not inbred, etc. The inverse of the numerator relationship matrix
$\mathbf{A}^{-1}$ can be easily obtained by using the Henderson's
rules \cite{henderson:84a}, but the whole coefficient matrix in equation
(\ref{eq13}) has to be explicitly inverted.

The amount of animals used in a typical analysis is around hundreds
of thousands and at least a matrix row per animal is needed so that
the rank of the coefficient matrix in Eq. (\ref{eq13}) is of that
order. This coefficient matrix is positive definite and the Gibbs
sampler based algorithm \cite{harville:99} can be easily implemented.
Nevertheless, the artificial selection based on the Henderson's mixed
model equations tend to select animals coming from a small number
of families and the percentage of inbreeding increases significantly
after a few generations of artificial selection. Wu and Schaeffer
\cite{wu:00} proposed an original method to select artificially the
populations keeping the genetic gain very close to the Henderson's
optimum but reducing significantly the rate of inbreeding. The numerator
relationship matrix has to be replaced by a customary asymmetric relationship
matrix $\widetilde{\mathbf{A}}$ , and they provide simple rules to
obtain $\widetilde{\mathbf{A}}^{-1}$. For each animal $i$ in the
pedigree, with sire $s$ and dam $d$,

\begin{enumerate}
\item Add $\left(\left(1-\lambda\right)\delta_{i}+\lambda\right)$ to the
$\left(i,i\right)$ position of $\widetilde{\mathbf{A}}^{-1}$, 
\item Add $-\left(1-\lambda\right)\delta_{i}/2$ to the $\left(i,s\right)$
and $\left(i,d\right)$ positions of $\widetilde{\mathbf{A}}^{-1}$, 
\item Add $-\delta_{i}/2$ to the $\left(s,i\right)$ and $\left(d,i\right)$
positions of $\widetilde{\mathbf{A}}^{-1}$, 
\item Add $-\delta_{i}/4$ to the $\left(s,s\right)$, $\left(d,d\right)$,
$\left(s,d\right)$ and $\left(d,s\right)$ positions of $\widetilde{\mathbf{A}}^{-1}$, 
\end{enumerate}
where $\delta_{i}=2$ when both parents of the $i^{th}$ animal are
known, $\delta_{i}=4/3$ when one parent is known and $\delta_{i}=1$
when both parents are unknown. $\lambda\in\left[0,1\right]$ is a
known coefficient which determine the weight of the family records
in each animal genetic evaluation, for instance, setting $\lambda=0$
correspond to the conventional Henderson's mixed model equations.
These rules result in an asymmetric coefficient matrix for $\lambda\neq0$,
non suitable for \emph{GS}. Both \emph{SE} and \emph{CC} can be used
to compute the inverse required in expression (\ref{eq13}) even in
cases where $\mathbf{A}$ is replaced by its non-symmetric counterpart
$\widetilde{\mathbf{A}}$. We will now compare the performance of
our algorithm against the \emph{SE} estimator using two cases of Wu
and Schaeffer's coefficient matrices as described in Table \ref{Tabla1}. 

\begin{table}

\caption{\label{Tabla1}Topics of the Wu and Schaeffer's coefficient matrix
in two cases of different size.}

\centering 

\begin{tabular}{|l|r|r|}
\hline 
Case&
 1&
 2\tabularnewline
\hline
Number of animals&
 50000&
 100000\tabularnewline
\hline
Number of herds&
 5000&
 10000\tabularnewline
\hline
Rank of the coefficient matrix&
 55000&
 110000\tabularnewline
\hline
Nonzero elements&
 424978&
 848982\tabularnewline
\hline
$\sigma_{e}^{2}/\sigma_{a}^{2}$&
 3.0&
 3.0\tabularnewline
\hline
$\lambda$&
 0.2&
 0.2  \tabularnewline
\hline
\end{tabular}

\par{}
\end{table}

The correlated chains sampling was implemented by setting the tolerance
for burn-in as $5\,10^{-5}$. After the burn-in period was finished,
realized values of $tr(\mathbf{z}^{\dagger}\mathbf{Qw})$ were obtained
for each cycle and averaged. Convergence after burn-in was checked
every $100^{th}$ iteration. The standard error of the final estimate
of $E\left[tr(\mathbf{z}^{\dagger}\mathbf{Qw})\right]$ was obtained
from the variance between realized values of $tr(\mathbf{z}^{\dagger}\mathbf{Qw})$
and an effective chain length as described in \cite{geyer:92}. Making
recourse of this effective length comes as a consequence of the correlated
nature of the realized values in the \emph{CC} algorithm. At any rate,
we tested the Geyer's method against an empirical standard error estimate
obtained by replicating one hundred times the whole analysis varying
the random seed number. The algorithm was assumed to reach the convergence
when the standard error yielded a relative error smaller than the
required tolerance of $5\,10^{-5}$.

\emph{SE} method was implemented by solving the linear systems by
both the iterative Gauss-Seidel algorithm and the bi-conjugate gradient
method. The latter resulted in a more efficient \emph{SE} algorithm
in terms of CPU time and it is the only one presented in this manuscript.

Table \ref{Tabla2} shows the results of both algorithms in cases
1 and 2. Relevant topics concerning the computing efficiency of the
\emph{CC} algorithm are the number of rounds computed to reach the
burn-in tolerance, the number of rounds required to reach the required
standard error of the final estimate, and the CPU time required to
compute both each burn-in cycle (four chains) and each after burn-in
cycle (two chains). Relevant topics for the \emph{SE} algorithm are
the average number of rounds required to solve each linear system
via the bi-conjugate gradient method (maximum change between successive
iterations was set to $5\,10^{-5}$), the number of linear systems
to be solved in order to reach the required standard error of the
final estimate and the CPU time per bi-conjugate gradient round of
iteration. All CPU times in Table \ref{Tabla2} are referred to the
time taken to complete a cycle in the \emph{CC} algorithm in the smallest
case.

Table \ref{Tabla2} shows also the agreement between the results provided
by both the \emph{CC} and the \emph{SE} algorithms. The \emph{CC}
algorithm is vastly more efficient in terms of the final CPU time
required for the computation of the inverse being around eight times
faster then its \emph{SE} counterpart. \emph{}

\begin{table}

\caption{\label{Tabla2}Results of \emph{CC} and \emph{SE} on the two Wu and
Schaeffer's matrices}

\centering 

\begin{tabular}{|l|r|r|l|r|r|}
\hline 
\multicolumn{3}{|l|}{\emph{CC} method}&
\multicolumn{3}{l|}{\emph{SE} method}\tabularnewline
\hline
&
 Case 1&
 Case 2&
&
 Case 1&
 Case 2\tabularnewline
\hline
$N$&
 103&
 104&
 Rounds per system&
 34.63&
 38.04\tabularnewline
\hline
$M-N$&
 39200&
 23800&
 Total rounds&
 285732&
 202448\tabularnewline
\hline
Eff. length size&
 15788&
 8163&
 Number of systems&
 8251&
 5322\tabularnewline
\hline
$E\left[tr(\mathbf{z}\prime'\mathbf{Qw})\right]$&
 10371&
 20738&
 $E\left[tr(\mathbf{z}\prime'\mathbf{Qw})\right]$&
 10371&
 20738\tabularnewline
\hline
$Var\left[tr(\mathbf{z}\prime'\mathbf{Qw})\right]$&
 3932&
 8123&
 $Var\left[tr(\mathbf{z}\prime'\mathbf{Qw})\right]$&
 2391&
 5301\tabularnewline
\hline
MC St. error&
 0.499&
 0.998&
 MC St. error&
 0.499&
 0.998\tabularnewline
\hline
Empirical St. error&
 0.502&
 1.047&
&
&
\tabularnewline
\hline
CPU time per burn-in cycle&
 1.97&
 3.99&
&
&
\tabularnewline
\hline
CPU time per cycle&
 1&
 2.03&
 CPU time per round&
 1.03&
 1.99\tabularnewline
\hline
CPU time per eff. cycle&
 2.48&
 5.92&
 CPU time per system&
 35.69&
 75.80\tabularnewline
\hline
Total CPU time&
 39403&
 48729&
 Total CPU time&
 294303&
 402872\tabularnewline
\hline
\end{tabular}

\par{}
\end{table}

\subsection{Example 2. The Dirac's free fermion complex non-hermitian matrix}

We now proceed further in assessing the performance of the \emph{CC}
algorithm by including complex matrix components as well as by strengthen
the number of non-zero elements (in the millions range). The tolerance
now set to $10^{-5}$ with respect to the absolute value rather than
for real and imaginary part independently. The chosen example is very
well representative of the computational demanding tasks typical of
the physical sciences. It belongs to the realm of elementary particle
physics known as Lattice Quantum Chromodynamics briefly described
in the following. 

In elementary particle physics the evolution of the particles is described
in terms of a field, $\psi\left(x\right)$, over the space-time ($x$,
is a four components vector, one representing time, the other three
the spatial coordinates) which governs the annihilation and creation
of particles. More specifically $\psi\left(x\right)$is an operator
whose action on the elements of its domain (a Hilbert space) represents
the annihilation of a particle at time $x_{0}$and spatial coordinates
($x_{1},\, x_{2},\, x_{3}$) . Conversely, the action of the adjoin
field $\psi^{\dagger}\left(x\right)$ (the hermitian conjugate in
a matrix representation of $\psi\left(x\right)$), describes the creation
of a particle at space-time coordinates $x$ . For a class of particles
called 1/2-spin Fermions, such as the electron, $\psi$is in turn
a four components object (each component being an operator) called
a spinor, an object without a classical analog (it is not a vector
since under a $2\pi$rotation changes its sing). Free evolution of
such particles, i.e., without interaction with other particles, is
governed by the Dirac's equation which in the so called Euclidean
representation reads\begin{equation}
L\,\psi\equiv(\partial_{\mu}\gamma^{\mu}+m)\psi=0,\label{eq:dirac}\end{equation}
 where $m$ is the mass of the particle, $\mu$ runs over the four
spatio-temporal coordinates, $\partial_{\mu}$ denotes derivative
along the $\mu$ coordinate, summation over repeated indexes is assumed
and $\gamma^{\mu}$ is a set of four non-hermitian matrices of dimension
four satisfying\[
\gamma^{\mu}\gamma^{\nu}+\gamma^{\nu}\gamma^{\mu}=2\delta_{\mu\nu}^{(4)},\]
 where $\delta_{\mu\nu}^{(4)}$denotes a four by four unit matrix
multiplied by the standard Kronecker delta. Alternatively the evolution
can be described through the associated Green function of Eq. (\ref{eq:dirac}),
that is, the solution of

\begin{equation}
L\, G(x,x')=\delta(x-x'),\label{eq:green}\end{equation}
 where $\delta(x-x')$represents the Dirac's delta. $G(x,x')$is known
also as a propagator of the field since given an initial configuration
for the field $\psi(x^{0})$, the field at any other space-time event
is obtained by\[
\psi(x)=\int G(x,x^{0})\psi(x^{0})dx^{0},\]
 that is, $G(x,x^{0})$represents the propagation of a Dirac's delta
signal from the event $x^{0}$ to the event $x$. In probabilistic
terms, it gives the probability of finding a particle at the space-time
event $x$ given that it was at the event $x^{0}$. Propagators are
of paramount importance in quantum field theory.

In general the evolution is far more complicated than that described
by (\ref{eq:dirac}) since interactions among several fields will
be effective. Thus, electrons, having electric charge will interact
between each other through the electromagnetic field (a gauge field
in the jargon of quantum field theory) or, in terms of particles,
by exchange of photons. Eq. (\ref{eq:dirac}) must then be supplemented
with both a free term for the evolution of photons and an interaction
term between photons and electrons. Within this frame (Quantum Electrodynamics)
the interaction is weak enough for a perturvative approach in many
interesting situations. However, there are other cases in which the
coupling is so strong that perturvative techniques are useless. The
paradigmatic case is the interaction of quarks (also fermions, like
the electrons) inside nucleons (protons and neutrons), this time by
exchange of particles called gluons describing an interaction known
as Quantum Chromodynamics (QCD). In such a case, recourse is made
to a discretization of the space-time (a space-time lattice) suitable
for a numerical solution of the problem. Under such discretization
Eq. (\ref{eq:green}) leads to a matrix equation of the form $\mathbf{L}\,\mathbf{G}=\mathbf{I}$,
so that the propagator is given by $\mathbf{L}^{-1}$. Interactions
make $\mathbf{L}$ depend on the value of the gluon field at each
spacetime event but still the discretized quark propagator is given
by a matrix inversion. However, an extra average over an ensemble
of different realizations of the fields is necessary. Altogether leads
to an extremely demanding computational task and Lattice QCD is responsible
of a mayor part of the CPU time consumed in Science (for a series
of reviews in the computational aspects of Lattice QCD see \cite{Cabibbo:99}
and in particular \cite{gusken:99}). Here we are only interested
in testing the CC algorithm so that we restrict ourselves to the free
case described by (\ref{eq:dirac} ). Although trivial from the physical
point of view, this simple case can be solved analytically \cite{Creutz:83}
and therefore the solution can be checked. The final result after
certain technical details reads (in a compact notation) \cite{Creutz:83},

\begin{equation}
L_{m\, n}=\delta_{m\, n}^{(4)}+K\sum_{\mu}\left(\left(1+\gamma^{\mu}\right)\delta_{m_{\nu}+\delta_{\mu\nu},\, n_{\nu}}^{(4)}+\left(1-\gamma^{\mu}\right)\delta_{m_{\nu}-\delta_{\mu\nu},\, n_{\nu}}^{(4)}\right),\label{eq:disDirac}\end{equation}
 where Latin indexes run over lattice points, Greek indexes over space-time
dimensions, $K$ is a constant approaching 1/8 in the continuum limit
and $\delta_{\mu\nu}$ is the standard Kronecker delta. In this notation
a Latin index, say $m$, labels a lattice point while the $\nu$ spatio-temporal
coordinate of such a point is denoted by $m_{\nu}$. Thus, in each
lattice point a four dimensional matrix acting on the spinor components
is defined.

Explicitly in terms of both the spatio-temporal coordinates (Latin
indexes) and spinor components (Greek indexes) $L$ reads

\begin{eqnarray}
L_{\mu\,\nu\: a\, b\, c\, d\: m\, n\, l\, k} & = & \delta_{\mu\nu}\delta_{a\, m}\delta_{b\, n}\delta_{c\, l}\delta_{d\, k}+\nonumber \\
 &  & K\left\{ \delta_{b\, n}\delta_{c\, l}\delta_{d\, k}\left[\delta_{a+1\, m}\left(\delta_{\mu\nu}+\gamma_{\mu\nu}^{1}\right)+\delta_{a-1\, m}\left(\delta_{\mu\nu}-\gamma_{\mu\nu}^{1}\right)\right]+\cdots\right\} .\label{eq:explicit}\end{eqnarray}
 In the next term in the sum $\gamma^{1}$ changes to $\gamma^{2}$,
$a$ interchanges with $b$and $m$ with $n$ and so for until the
indexes are exhausted. The $\gamma$'s matrices are defined as\[
\mathbf{\gamma}^{i}=\left(\begin{array}{cc}
0 & \sigma^{i}\\
\sigma^{i} & 0\end{array}\right),\: i=1,\,2,\,3\]
 and\[
\mathbf{\gamma}^{4}=\left(\begin{array}{cc}
0 & 1\\
1 & 0\end{array}\right)\]
 where the $\mathbf{\sigma}$'s, the $1$ and the $0$ must be understood
as two by two matrices. The $\sigma$'s are the show called Pauli
matrices given by\begin{eqnarray*}
\mathbf{\sigma}^{1} & = & \left(\begin{array}{cc}
0 & 1\\
1 & 0\end{array}\right),\end{eqnarray*}

\begin{eqnarray*}
\mathbf{\sigma}^{2} & = & \left(\begin{array}{cc}
0 & -i\\
-i & 0\end{array}\right),\end{eqnarray*}

\begin{eqnarray*}
\sigma^{3} & = & \left(\begin{array}{cc}
1 & 0\\
0 & -1\end{array}\right),\end{eqnarray*}
 with $i$ representing the imaginary unit. Eq. (\ref{eq:explicit})
is useful to implement $L$ in a program but to apply the \emph{CC}
algorithm as described in in section 2.4, a one to one mapping of
the ten indexes of $L$ to only two is also needed. We use the following
one usual in QCD\begin{eqnarray*}
m' & = & 1+a+N_{1}\left(b+N_{2}\left(c+N_{3}\left(d+N_{0}\mu\right)\right)\right)\\
n' & = & 1+m+N_{1}\left(n+N_{2}\left(l+N_{3}\left(k+N_{0}\nu\right)\right)\right),\end{eqnarray*}
Here $N_{1}$, $N_{2}$, $N_{3}$ and $N_{0}$ are the number of discrete
points in each spatio-temporal coordinate of the lattice (as above,
$0$ corresponds to the time coordinate). The rank of the corresponding
matrix is then $4\, N_{0}N_{1}N_{2}N_{3}$ while expression (\ref{eq:explicit})
yields fourteen non zero elements per row so that the total number
of non zero elements is $56\, N_{0}N_{1}N_{2}N_{3}$.

Again two cases were tested. The parameters are given in table \ref{Tabla3}
and corresponds to two different spatio-temporal lattice sizes in
both cases with the same number of discrete values for the four coordinates.
The value of $K$ was arbitrarily chosen to 0.1. 

\begin{table}[H]

\caption{\label{Tabla3}Topics of the Dirac's free fermion matrix in two cases
with different lattice sizes}

\centering 

\begin{tabular}{|l|r|r|}
\hline 
Case&
 1&
 2\tabularnewline
\hline
$N_{1}$, $N_{2}$, $N_{3}$, $N_{0}$&
 18&
 20\tabularnewline
\hline
Rank of $\mathbf{C}$&
 419904&
 640000\tabularnewline
\hline
Nonzero elements&
 5878656&
 8960000\tabularnewline
\hline
$K$&
 0.1&
 0.1  \tabularnewline
\hline
\end{tabular}

\par{}
\end{table}

\begin{table}[H]

\caption{\label{Tabla4}Results of \emph{CC} and \emph{SE} on Dirac's matrices
(first case). }

\centering

\begin{tabular}{|l|r|l|r|}
\hline 
\multicolumn{2}{|l|}{\emph{CC} method}&
\multicolumn{2}{l|}{\emph{SE} method}\tabularnewline
\hline
$N$&
 18&
 Rounds per system&
 55.03\tabularnewline
\hline
$M-N$&
 10832&
 Total rounds&
 219167\tabularnewline
\hline
Eff. length size&
 10805.0&
 Number of systems&
 3983\tabularnewline
\hline
$E\left[tr(\mathbf{z}^{\dagger}\mathbf{Qw})\right]$ exact value&
413007.84+0i&
&
\tabularnewline
\hline
$E\left[tr(\mathbf{z}^{\dagger}\mathbf{Qw})\right]$&
 413004.47-1.87i&
 $E\left[tr(\mathbf{z}^{\dagger}\mathbf{Qw})\right]$&
 413005.08-1.98i\tabularnewline
\hline
$Var\left[tr(\mathbf{z}^{\dagger}\mathbf{Qw})\right]$&
 184130.43&
 $Var\left[tr(\mathbf{z}^{\dagger}\mathbf{Qw})\right]$&
 67827.29\tabularnewline
\hline
MC St. error&
 4.128&
 MC St. error&
 4.130\tabularnewline
\hline
CPU time per burn-in cycle&
 1.43&
&
\tabularnewline
\hline
CPU time per cycle&
 0.4746&
 CPU time per round&
 0.40\tabularnewline
\hline
CPU time per eff. cycle&
 1.00&
 CPU time per system&
 21.89\tabularnewline
\hline
Total CPU time&
 10859&
 Total CPU time&
87167 \tabularnewline
\hline
\end{tabular}

\par{}
\end{table}

\begin{table}[H]

\caption{\label{cap:Results-of-CC}Results of \emph{CC} and \emph{SE} on Dirac's
matrices (second case). }

\centering 

\begin{tabular}{|l|r|l|r|}
\hline 
\multicolumn{2}{|l|}{\emph{CC} method}&
\multicolumn{2}{l|}{\emph{SE} method}\tabularnewline
\hline
$N$&
 18&
 Rounds per system&
 56.08\tabularnewline
\hline
$M-N$&
 6782&
 Total rounds&
 146379\tabularnewline
\hline
Eff. length size&
 6848.8&
 Number of systems&
 2610\tabularnewline
\hline
$E\left[tr(\mathbf{z}^{\dagger}\mathbf{Qw})\right]$ exact value&
629489.14+0i&
&
\tabularnewline
\hline
$E\left[tr(\mathbf{z}^{\dagger}\mathbf{Qw})\right]$&
 629480.89-0.53i&
 $E\left[tr(\mathbf{z}^{\dagger}\mathbf{Qw})\right]$&
 629482.78-0.1i\tabularnewline
\hline
$Var\left[tr(\mathbf{z}^{\dagger}\mathbf{Qw})\right]$&
 270373.73&
 $Var\left[tr(\mathbf{z}^{\dagger}\mathbf{Qw})\right]$&
 103395.82\tabularnewline
\hline
MC St. error&
 6.283&
 MC St. error&
 6.294\tabularnewline
\hline
CPU time per burn-in cycle&
2.26&
&
\tabularnewline
\hline
CPU time per cycle&
 1.54&
 CPU time per round&
0.61\tabularnewline
\hline
CPU time per eff. cycle&
1.53&
 CPU time per system&
33.99\tabularnewline
\hline
Total CPU time&
 10503&
 Total CPU time&
88721 \tabularnewline
\hline
\end{tabular}

\par{}
\end{table}

From inspection of tables \ref{Tabla4} and \ref{cap:Results-of-CC}
it is clear the advantage of the \emph{CC} algorithm whose performance
again is around eight times higher than that of the \emph{SE}. As
explained above, this time it is possible to check the results against
exact deterministic calculations. As expected they coincide with the
estimated values within the imposed tolerance.

\section{Discussion and Conclusion}

Once the far superior efficiency of the \emph{CC} algorithm has been
demonstrated some comments about its applicability seems in order
but first a note about the \emph{GS} . Notice that the \emph{GS} as
implemented in (\ref{eq2}) is just the particular case of \emph{CC}
corresponding to hermitian matrices since in this case $\mathbf{z^{\dagger}=w}$.
Therefore its convergence is determined by $sp(\mathbf{T})<1$ and
the no singularity of $\mathbf{D}$. These conditions do not implies
$\mathbf{C}$ being positive definite. For instance, in the trivial
case of $\mathbf{C}$ being diagonal, $sp(\mathbf{T})=0$, and Eq.
(\ref{eq2}) reduces to $z_{i}^{\left(k\right)}=\phi_{i}^{\left(k\right)}/\sqrt{c_{ii}}$
which, although implies imaginary values of $z_{i}^{\left(k\right)}$
works perfectly. 

In the general case when the convergence criteria are not satisfied
at least two alternatives exits. One is to rewrite the algorithm with
a different partition than that shown here. For instance, suppose
that $\mathbf{D}$ is singular. Then, a possibility is to use a partition
with relatively small non singular diagonal blocks surrounding the
zeros of $\mathbf{D}$, amenable of been inverted deterministically
with the memory recourses available. In fact, in terms of time efficiency,
there will be partitions more efficient than the one used here but,
obviously, without the appealing implementation simplicity of Eqs.
(\ref{eq4}) and (\ref{eq5}). As a rule of the thumb, the convergence
will be guaranteed for a sufficiently {}``heavy'' $\mathbf{D}$.
A new partition in the way just described could solve problems with
small elements in the original $\mathbf{D}$ not only with zeros. 

Another simpler possibility is just to reorder the rows and the columns
of $\mathbf{C}$ trying to locate in the diagonal large enough elements.
In general, reordering which implies a low time penalty, could be
advantageous to improve the convergence of the method.

In summary, we have presented an efficient stochastic algorithm based
in correlated Markov chains suitable for the inversion of very large
matrix whenever the memory recourses are not enough for the application
of the standard deterministic methods. The efficiency of the algorithm
has been tested in a couple of numerical examples rendering a dramatic
improving of eight times faster runs with respect to the best stochastic
method known by the authors. The necessary and sufficient conditions
for the convergence of the algorithm have been also given.

\end{document}